# Flexible daytime radiative cooling enhanced by enabling three-phase composites with scattering interfaces between silica-microspheres and hierarchical porous coatings

*Hongchen Ma[a#], Liang Wang[b#], Shuliang Dou[c#], Haipeng Zhao[a], Min Huang[a], Zewen Xu[d], Xinyuan Zhang[a], Xiudong Xu[a], Aiqin Zhang[d], Huiyu Yue[d], Ghulam Ali[d], Caihua Zhang[d,e], Wen-Ying Zhou[e], Yao Li[c], Yaohui Zhan[a,f\*] and Cheng Huang[d,f\*]*

[a]*School of Optoelectronic Science and Engineering & Key Lab of Advanced Optical Manufacturing Technologies of Jiangsu Province, Soochow University, Suzhou 215006, China; E-mail: yhzhan@suda.edu.cn*
[b]*State Key Laboratory of High Performance Ceramics and Superfine Microstructure, Shanghai Institute of Ceramics, Chinese Academy of Sciences, Shanghai 200050, China.*
[c]*National Key Laboratory of Science and Technology on Advanced Composites, Harbin Institute of Technology, Harbin 150001, China.*
[d]*Soochow Institute for Energy and Materials InnovationS, College of Energy, Key Laboratory of Advanced Carbon Materials and Wearable Energy Technologies of Jiangsu Province, Soochow University, Suzhou 215006, China; E-mail: chengh@suda.edu.cn*
[e]*School of Chemistry and Chemical Engineering, Xi'an University of Science and Technology, Xi'an 710054, China.*
[f]*Light Industry Institute of Electrochemical Power Sources, Zhangjiagang 215600, China.*
[#]*The authors contributed equally.*

**ABSTRACT**

Daytime radiative cooling has attracted considerable attention recently due to its tremendous potential for passively exploiting the coldness of deep-sky as clean and renewable energy. Many advanced materials with novel photonic micro-nanostructures have already been developed to enable highly efficient daytime radiative coolers, among which the flexible hierarchical porous coatings (HPCs) are a more distinguished category. However, it is still hard to precisely control the size distribution of the randomized pores within the HPCs, usually resulting in a deficient solar reflection at the near-infrared optical regime under diverse fabrication conditions of the coatings. We report here a three-phase (i.e., air pore-phase, microsphere-phase and polymer-phase) self-assembled hybrid porous composite coating which dramatically increases the average solar reflectance and yields a remarkable temperature drop of ~10℃ and ~30℃ compared to the ambient circumstance and black paint, respectively, according to the rooftop measurements. Mie theory and Monte Carlo simulations reveal the origin of the low reflectivity of as-prepared two-phase porous HPCs, and the optical cooling improvement of the three-phase porous composite coatings is attributed to the newly generated interfaces possessing the high scattering efficiency between the hierarchical pores and silica microspheres hybridized with appropriate mass fractions. As a result, the hybrid porous composite approach enhances the whole performance of the coatings, which provides a promising alternative to the flexible daytime radiative cooler.

# ◼ INTRODUCTION

Daytime radiative cooling has recently aroused intense interest because it can dissipate terrestrial heat to deep space and yield a below-ambient temperature without any external energy input, even under direct sunlight[1–7]. The unique nature of radiative cooling has enabled promising applications in various fields such as energy-efficient buildings[8–10], artificial wood[11], solar cells[12–14], water condensation[15–17], and thermoelectric devices[18]. In most of these application scenarios, daytime radiative cooling (DRC) is urgently needed rather than nocturnal cooling, yet it is also more challenging since the solar radiation generally far exceeds thermal emission during the day. Many advanced photonic structures have already been developed towards maximizing both the solar reflectivity and the thermal emissivity simultaneously, such as multilayered films[19–22], nanoparticle-based composite materials[23–30], and porous media coatings[1–3,31–33]. For example, Y. Zhai et al. fabricated a kind of non-porous two-phase (i.e., microsphere-phase and polymer-phase) hybrid metamaterials in a roll-to-roll manner, which exhibits a strong emissivity (~ 0.93) in the mid-infrared regime but a high transmittance in the visible range; with the silver coating deposited as a back reflector, the solar reflectivity can be increased to ~96% and thus yielding a cooling power of 93W/$m^2$ at noontime[7]. Although silver and other metal mirrors are widely employed in the radiative coolers, it is generally expensive and prone to cause light pollution in practical use due to the specular reflection. Furthermore, the metal reflector fails to work near the ultraviolet from the strong intrinsic absorption[34]. Therefore J. Mandal et al. designed an alternative to two-phase (i.e., air-phase and polymer-phase) hierarchically porous poly(vinylidene fluoride-co-hexafluoropropene) [P(VdF-HFP)] coating with well-controlled pore-size distributions, exhibiting a sub-ambient temperature drop of 6°C and a cooling power of 96 W/m$^2$ under direct sunlight[1]. The two-phase hierarchical porous coatings (HPCs) with light scattering air porosity as virtual microspheres embedded in the coatings instead of the real microspheres by the pore-making technology reflect almost all the solar energy by the diffuse scattering of hierarchically pores, even without metal reflectors. Moreover, the wet solution-based approach shows a paint-like simplicity when applied to the diverse surfaces such as the exterior wall of buildings. And the cheap, flexible, and scalable coatings from the HPCs might facilitate the large-scale application of radiative cooling[35–38].

Although the two-phase porous HPCs exhibit some outstanding performance on solar scattering, yet it is still hard to precisely control the pore-size distribution during the formation of air pores. The air porous structures and porosity distribution significantly influence the solar reflectivity. In general, as a practical and urgent materials issues, an insufficient solar reflection can always be measured from the as-prepared porous coatings without deep optimization, especially in the near-infrared wavelengths. Therefore the microstructures of the HPCs would be reshaped with controllable pore-size distribution to significantly improve the solar reflectivity, alternatively, reconstructed to greatly enhance the microstructural scattering via effective measures.

Herein, we report an architectural design and fabrication of the three-phase self-assembled hybrid porous composite coatings possessing an appropriate amount of silica microspheres embedded within hierarchical porous heterostructures, enabling additionally created microsphere/air interfaces to dramatically enhance the Mie scattering and flexible daytime radiative cooling. It is also mentioned that this three-phase porous composite approach is some different from the previous two-phase porous or non-porous composite strategies for increasing solar reflectivity[22,27,29,39–43], In general the phase inversion fabrication method results in two-phase porous composite coatings, while direct solution cast methods produce the two-phase non-porous composites. Here the modified phase inversion fabrication method is introduced to the formation of three-phase porous composite coatings by controlling the mass fractions of silica microspheres. With an appropriate amount of silica microspheres the composite maintains the architecture of hierarchical pores, while it is hard to apply to the porous coatings when hybridizing with a higher amount of microspheres and/or nanospheres in homogeneous or heterostructured media since the densely distributed microspheres and/or nanospheres might be easy to occupy or destroy the hierarchical porous structure. By combining the air pores and silica microspheres with the polymer phase, a three-phase hybrid porous composite with high interface optical scattering and cooling performance is demonstrated. In this three-phase hybrid porous composite with moderate silica microspheres, the averaged solar reflectivity and the theoretical cooling power can reach 0.94 and 130 W/m$^2$, respectively. As a result, the rooftop temperature test shows a sub-ambient temperature drop of approximately 10℃ under intense solar irradiance. For this purpose of solar reflectivity enhancement, the influence of the added microsphere phase on the microstructures and optical scattering interfaces as well as cooling performance will also be systematically investigated.

## ◼ RESULTS AND DISCUSSION

The design concept for three-phase self-assembled hybrid porous composite (i.e., air pore-phase, silica microsphere-phase, and polymer-phase) and enhanced optical scattering interfaces is schematically shown in Figure 1. As illustrated in Figure 1(a), the silica microspheres (in red) are randomly scattered in the P(VdF-HFP) matrix (in gray) with hierarchical pores (in light blue). Light background color represents the P(VdF-HFP) matrix to show the morphology of pores and microspheres clearer. The enlarged inset shows the representative three-phase material interfaces (i.e., silica/air, silica/polymer, and air/polymer) as the silica microspheres are introduced in the porous polymers, which collectively facilitate Mie scattering occurring to form strong diffuse reflection. The silica microspheres are chosen due to their wide availability and losslessness in the solar spectral range. The combination of silica microspheres and porous polymers in the composite design is expected to manifest its potential optical advantage in three aspects. Firstly, as shown in Figure 1(b), the refractive index of silica is higher than that of P(VdF-HFP), leading to an approximately 10% enhancement of index contrast to the air. Secondly, the microscale spheres might produce artificial microstructures to make up for the insufficient existence of micropores if that is the case. Thirdly, the silica microspheres can broaden the phonon resonances in glass-polymer metamaterials[7], which is expected to be preserved in the modified system. Overall, the design of embedding silica microspheres is anticipated to improve the solar reflectivity of non-optimized HPCs by reshaping the microscopic structures; In the meanwhile, the other unique performances of HPCs, i.e., ultra-low solar absorption and near-perfect infrared emissivity, are expected to be retained intact as well.

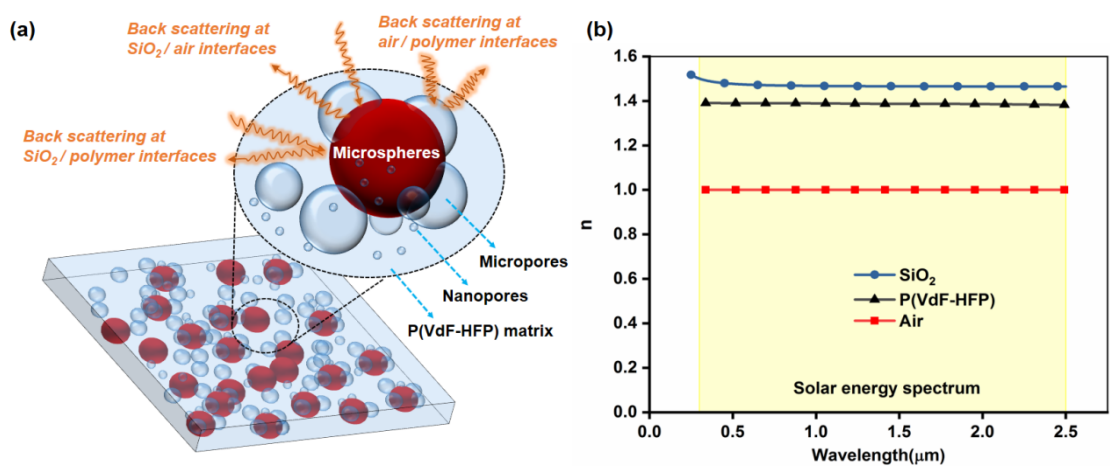

**Figure 1.** (a) Conceptual sketch for three-phase self-assembled hybrid porous composite (i.e., air pore-phase, silica microsphere-phase, and polymer-phase) and interfaces for enhancing the optical scattering of porous polymer coatings by embedding the silica microspheres therein. The light gray color of the background represents the P(VdF-HFP) matrix for highlighting the morphology of hierarchical pores (in light blue) and silica spheres (in dark red). The introduction of silica microspheres generates additional silica/air and silica/polymer interfaces within the self-assembled hybrid porous composite, which facilitate Mie scatterings contributing to better total solar reflection. (b) Refractive index contrast comparison of silica and P(VdF-HFP) materials with respect to air within the solar spectrum, indicating the possibility of stronger scattering occurring at the interface between silica and air media.

Figure 2(a) shows the preparation procedure for three-phase self-assembled hybrid porous composite coatings, based on the modified phase inversion method for fabricating hierarchically porous coatings[1] (See Method section for the detailed information). In general, the coating changes from transparent to opaque and eventually diffusing into the whole white film when the solvents evaporate (see Figure S1 for the optical images). It was found that the coatings exhibit the local wrinkling instability enhancement as the coating thickness increases (as shown in Figure S1). This pronounced wrinkling instability might be attributed to the non-uniform local inner stress distribution of asymmetric porous membranes, especially for the formation of thick membranes by the solvent evaporation. Therefore, the thinner coatings with a moderate thickness of 250 μm are preferred for a flat appearance, which is uniformized by using a home-made template (as shown in Figure S2). Figure 2(b) and (c) show the top-view and cross-sectional micrographs of porous two-phase HPCs (i.e., air pore-phase and polymer-phase), respectively. As shown in Figure 2(b), the coating surface presents a mesh-like morphology, which seems to be formed by the vapor phase escaping out of the surface. Figure 2(c) shows the typical porous characteristic that the nanoscale and microscale hierarchical pores are distributed in a random disorder state. The formation and morphologies of porous three-phase composites (i.e., air pore-phase, microsphere-phase and polymer-phase) were investigated by hybridizing silica microspheres with different mass fractions (Figure 2 and S3&4). Figure 2 (e-g) show typical microstructural morphologies of porous three-phase composite interfaces as the silica microspheres with a mass fraction of 9% are self-assembled into the three-phase composites. The surface morphology [as shown in Figure 2(d) and (e)] and the hierarchical feature of the polymer matrix [as shown in Figure (f) and (g)] indicate that the addition of moderate silica microspheres does not significantly influence the formation of hierarchical pores of air and polymer matrix. Furthermore Figure 2(f) and (g) show the typical microspheres being surrounded by the nanopores and micropores, respectively, which indicates that more air-microsphere and microhere-polymer interfaces occur as an appropriate amount of silica microspheres are added into the composites. However, as shown in Figure 2(h) and (i), as the fraction of microspheres is further increased (e.g., 15%) the mesh-like morphology and the micropores are almost destroyed since the excess microspheres greatly squeeze the amount of liquid phase in the solution and consequently impede the formation of the vapor phase, both in quantity and in velocity. Therefore as the vapor formation is slow enough, the microspheres are well-dispersed and can also adaptively fill the void space toward forming new non-porous two-phase composite (i.e., microsphere-phase and polymer-phase).

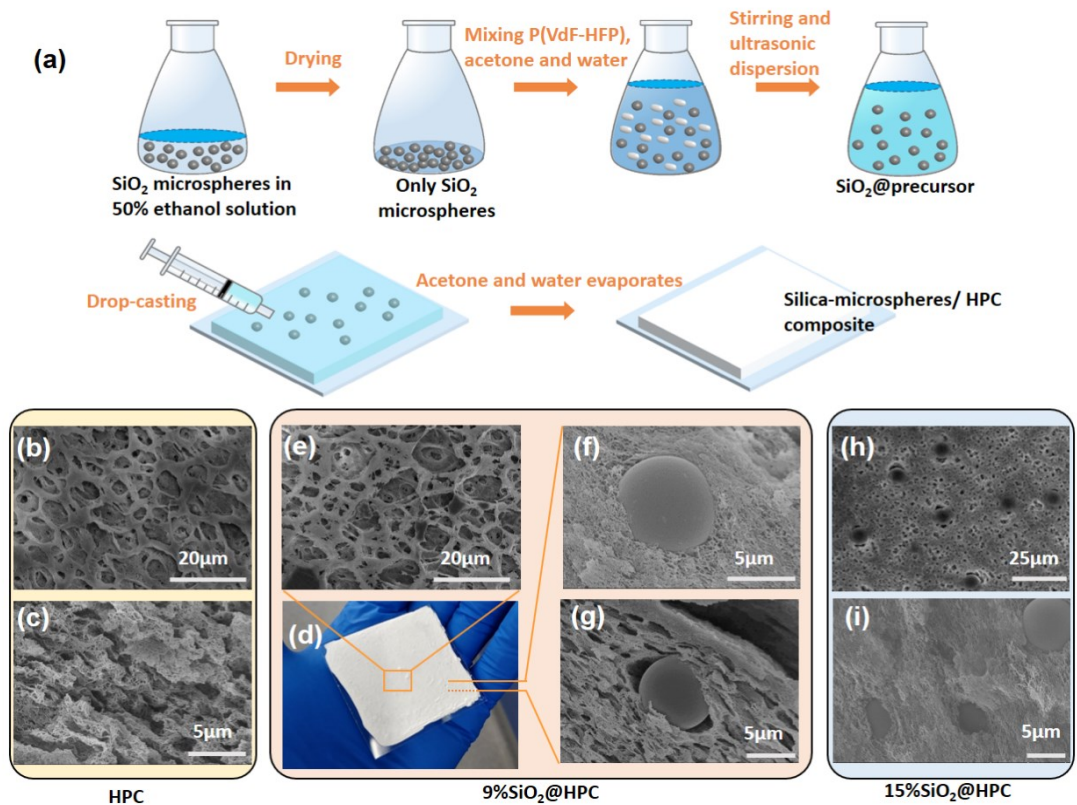

**Figure 2.** Preparation procedure and typical microscopic morphology. (a) Schematic of the detailed process for three-phase self-assembled hybrid porous composite coatings. (b) Top-view and (c) cross-sectional SEM images for as-prepared porous two-phase HPCs. (d) Photograph and (e-g) micrographs showing in (e) top-view and (f, g) cross-sectional views for the porous three-phase HPCs with 9% silica microspheres. The cross-sections in Figure (f) and (g) show the typical microspheres being surrounded by the nanopores and micropores, respectively. (h) Top-view and (i) cross-sectional views of the composite with 15% silica microspheres, trending to form a new non-porous two-phase composite (i.e., microsphere-phase and polymer-phase).

**Optical performance and power evaluation.** Figure 3(a) shows the experimental data of the reflectivity spectra in the range where solar irradiation dominates and the emissivity spectra in the mid-infrared range where the atmospheric window opens, for the composite coatings with a varying fraction (0% - 15%) of silica microspheres, respectively. As for the two-phase HPC sample without the presence of silica microspheres, the reflectivity is nearly perfect in the ultraviolet range (e.g., ~100% at 300 nm); however, it drops gradually as the wavelength increases, falling to approximately 90% at 600 nm and 70% at 1250 nm. Although the solar irradiance is relatively weak in the long-wavelength range, the power density integrated across the broadband spectrum is also unneglectable for the optimal radiative cooling performance.

By hybridizing HPCs with silica microspheres of different mass fractions, the solar reflectivity is indeed increased as the whole, except for a slight decline in a small portion of the ultraviolet band. As the mass fraction is 3%, the reflectivity spectrum exhibits a significant increase across the visible to the near-infrared range (i.e., by 6% and 14% at 600 nm and 1250 nm, respectively). As the mass fraction increases by 3% gradually, the near-infrared reflectivity monotonously grows at first and then declines dramatically. As shown on the right side of Figure 3(a), the emissivity (or 1-emissivity) maintains high (or low) across the atmospheric window, indicating that the mass fractions of silica microspheres almost do not influence the mid-infrared atmospheric window. To quantitatively evaluate the overall performance, the solar reflectivity and relevant power densities are calculated (see Method for the detailed physical expression and formula) and presented in Figure 3(b) and (c) for the different coatings. As shown in Figure 3(b), the HPC with an absence of microspheres exhibits a relatively low $R_{solar}$ of ~82%, which is improved by 8.1%, 11.4%, 14.5%, and 14.8% as introducing the silica microspheres with mass fractions of 3%, 6%, 9%, 12%, respectively. The increment in reflectivity shows a positive correlation to the mass fraction due to the superposition of Mie scattering, which becomes saturated at a fraction of 12%. As the mass fraction is further increased to 15%, the solar reflectivity drops back to ~86%, which might be ascribed to the destruction of beneficial micropores by excessive microspheres [as shown in Figure 2(h-i)].

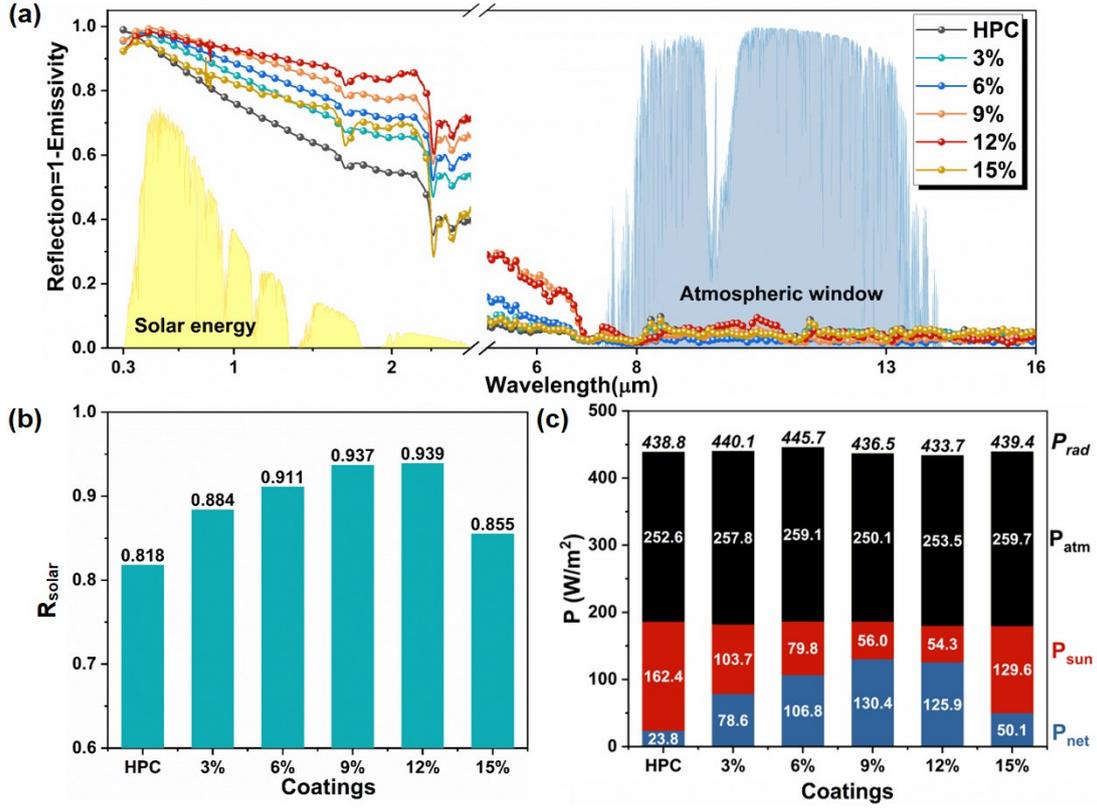

**Figure 3.** Optical property and cooling power evaluation. (a) Measured reflection and emission spectra of HPCs with a varying fraction (0% - 15%) of silica microspheres. The normalized ASTM G173 Global solar spectrum and atmospheric transparency window are included for illustrating spectral selectivity. (b) Calculated solar reflectances ($R_{solar}$) weighted by the standard solar irradiance of air mass 1.5 ($I_{AM1.5}$). (c) Power densities of $P_{rad}$, $P_{atm}$, $P_{sun}$, and $P_{net}$ for different coatings according to the measured spectra.

To take the mid-infrared emissivity into account, the power density values are calculated based on the well-developed thermal balance theory[19–21]. The reflectivity and emissivity are assumed to be angle independent for simplicity, which is reasonable for such a non-angle selective system with random porous microstructures[1]. In addition, the cooler temperature within the calculations is assumed to be the same as the ambient temperature, which is adopted as 313 K according to the following measurement. As shown in Figure 3(c), the sample of 6wt% of silica microspheres has the greatest $P_{rad}$ of ~446 W/m², consistent with the highest emissivity shown in Figure 3(a), especially in the atmospheric window. However, $P_{atm}$ of this sample is also higher than most of others, which cancels out the contribution of $P_{rad}$ to a certain degree. In contrast, $P_{atm}$ of the sample with 9wt% of silica microspheres is lowest (~250 W/m²), which is consistent with its lowest emissivity beyond 8-14 μm where the atmosphere is opaque for thermal irradiation. While both $P_{atm}$ and $P_{atm}$ have inconspicuous differences among samples, $P_{sun}$ plays a dominant role in yielding the final ranking of $P_{net}$. Based on the comprehensive comparison, the sample with 9wt% of silica microspheres wins first place with a $P_{net}$ of ~130 W/m², slightly surpassing that of the sample with 12wt% of silica microspheres due to a larger ($P_{rad}-P_{atm}$) value.

**Rooftop temperature measurements.** The rooftop parameter measurements were performed at Soochow University (N31°18′, E120°34′) on some sunny days in June of 2020. As shown in Figure 4, the test apparatus is developed by a foam box that is sealed with aluminum foil on the outer sides for better thermal insulation. A low-density polyethylene film covers the upper opening of the foam box, forming an air chamber to suppress convection and conduction. All the coatings are placed in the same apparatus to ensure an identical ambient condition and avoid the interference of different non-physical factors. The glass with black paint is utilized as a substrate to eliminate the back-reflection from the underlying base. The sample temperatures are characterized by the thermocouples placed between the samples and the black paint, while the temperatures of the ambient circumstance and the black paint are monitored by the thermocouples suspended in the air chamber and stuck onto the surface of the black paint, respectively. The inset on the right side of Figure 4(a) schematically shows the cross-section of the testing setup. The temperature data was stored every minute on a USB flash drive by the multichannel temperature logger. The humidity and solar irradiance are also recorded by the hygrometer and illuminometer at the same time.

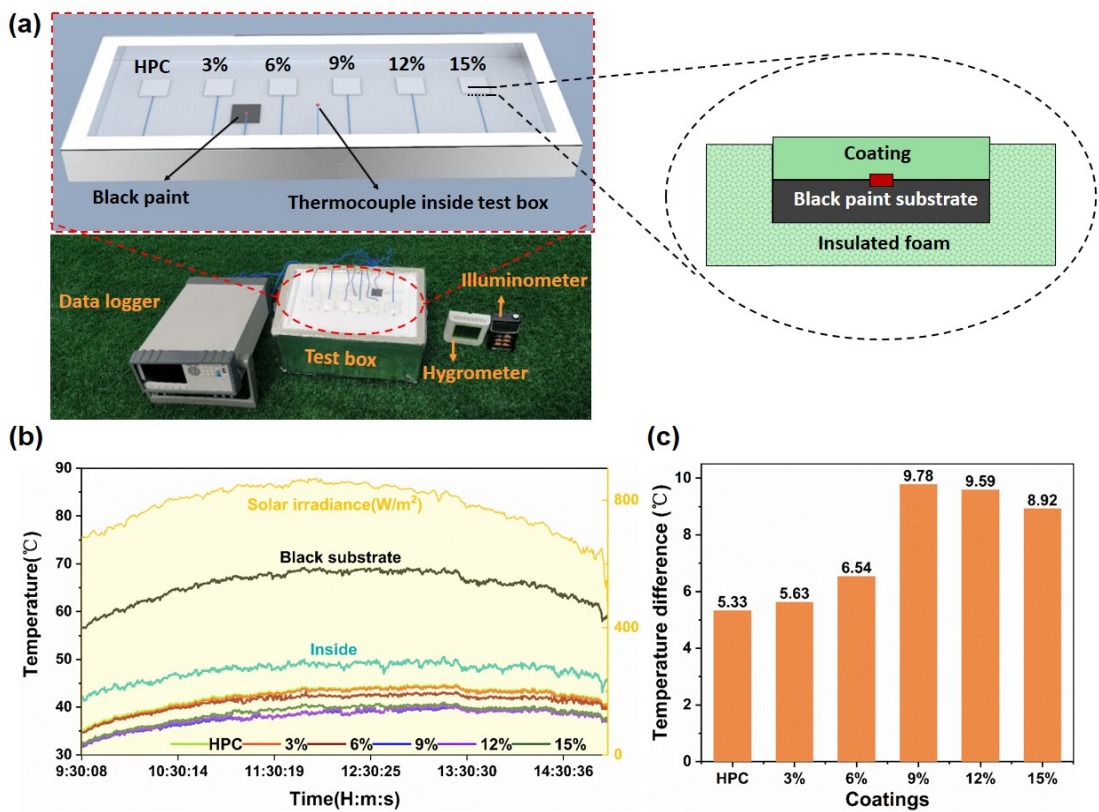

**Figure 4.** Daytime radiative cooling performance. (a) Schematic of the setup, sample holder and instrumentation for the rooftop test under sunlight. (b) Temperature tracking of the coatings, the inside cavity, and the black substrate. The monitored solar irradiance is included for providing primary meteorological information. (c) The average value of the temperature drops during the daytime for the coating samples compared to the inside air circumstance.

Figure 4(b) shows the daytime temperature measurement with the accompanied solar irradiance (see Figure S5 for the long-time continual rooftop test with the associated meteorological information). The solar irradiance peaks at noon with a power density exceeding 860 W/m$^2$, when the ambient temperature and the black paint temperature rise to approximately 50℃ and 70℃, respectively. Despite so, all the samples show temperatures beneath the ambient with the amount of temperature drop proportional to the solar reflectivity as discussed above. The cooling effect of the porous three-phase composite coatings with 9wt% and 12wt% of silica microspheres is more significant than that of the others, which exhibit a temperature no more than 40℃ under the strongest solar irradiance. The temperature of porous two-phase HPCs without microspheres is highest but almost overlapped by that of the three-phase composite coating with 3wt% of silica microspheres. To display the cooling performance clearer, the averaged temperature drops are summarized and derived in Figure 4(c) for all the coatings. As shown in Figure 4(c), the averaged temperature drop is gradually increased by introducing moderate silica microspheres, reaching an extreme value of 9.78℃ at a 9% of mass fraction followed by 9.59℃ at 12wt%, which are mainly coincident well with the calculated net cooling powers.

**Microscopic multiphase physics exploration.** Monte Carlo (MC) simulations were performed to reveal the physics underlying the experiments quantitatively. Figure 5(a) shows the reflectivity spectrum of the reference porous two-phase HPCs (blue line) obtained by MC calculations, which coincides nicely with the experiment (red line); the inset shows the corresponding pore-size distribution inversely obtained via spectral fitting. As shown in the inset, the nanopores are predominant in a relative ratio of 52%, whereas the micropores of ~5μm in diameter account for only 19% which are believed to possess an efficient sunlight scattering. The reduced reflection in the near-infrared regime might mainly originate in the insufficiency of micropores, while the abundant nanopores ensure the near-unitary reflectivity in the short-wavelength range [also see Figure S6 for the influence of the ratio of the micropores and/or nanopores for the porous two-phase composite coatings on the total reflectivity spectra]. While hybridizing with HPCs with silica microspheres, the microspheres might be surrounded by either pores or polymers within porous three-phase composite coatings from a probability perspective. Therefore, the main or key contributor to the enhancement of solar reflectivity was investigated. As shown in Figure 5(b), the scattering peaks occur only in the short-wavelengths (e.g., < 0.85 μm) for the silica microspheres in polymers, while they occur in both the short- and long-wavelength range for the same microspheres in air. Based on the microscopic scattering response, the macroscopic reflectivity is obtained using MC simulations for the extreme cases that only the polymer (denoted by microsphere@polymer) or the air (denoted by microsphere@air) surrounds the microsphere. The coating thickness within the calculation is exemplified as 50 μm considering a porosity of 20% as previously fitted for Figure 4(a), without loss of generality for qualitative exploration. As shown in Figure 5(c), the contribution to reflection keeps at a fairly low level for the case of microsphere@polymer, despite slightly increasing as the volume fraction increases. The volume fraction is employed here mainly due to the high-contrast material densities of two host media. By comparing

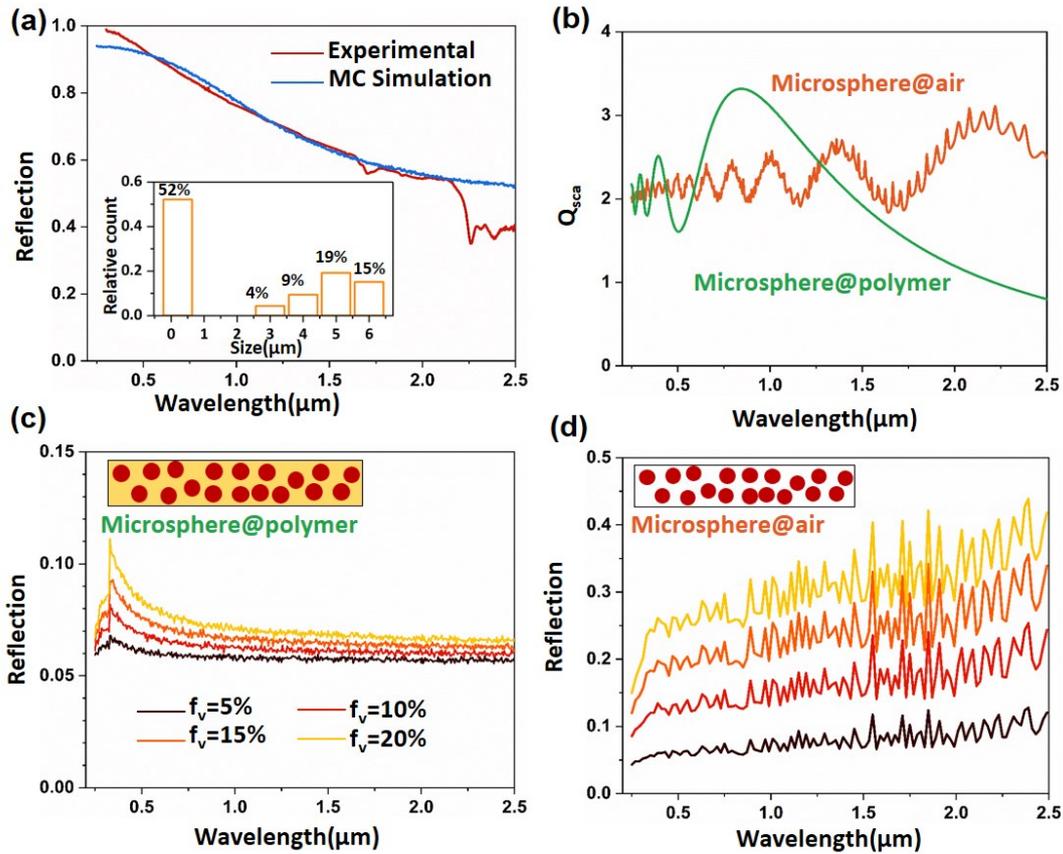

**Figure 5.** Underlying physics revealed by Mie scattering based Monte Carlo simulations. (a) Simulated (Blue) and measured (red) macroscopic reflectivity spectra for the as-prepared porous two-phase HPCs. Inset: the modeled pore-size distribution obtained by fitting the measured reflectivity spectrum. (b) Scattering efficiencies of an individual silica microsphere immersed in the non-porous polymer P(VdF-HFP) and air media, respectively. (c) and (d) Reflectivity spectra for the silica microsphere of different volume fractions of $f_v$ embedded in the (c) non-porous polymer and (d) air media.

Figure 4(b) with 4(c), it is found that although the scattering peaks are not precisely maintained at the same frequency of the reflection spectra, the tendency is similar that both the stronger scattering and reflection occur at the short-wavelength range. Figure 5(d) shows the case of the microsphere@air. The coating reflectivity is similar to that of the microsphere@polymer counterpart when the volume fraction is small (e.g., 5%); as the volume fraction increases gradually, the reflectivity is improved significantly, especially at the near-infrared range. Because there are two kinds of scatters, the total reflection spectra in the realistic porous three-phase hybridized system are challenging to be obtained by MC simulations directly; however, they can be imprecisely approximated by adding each reflectivity spectrum in Figure 5(d) with that of bare HPCs in Figure 5(a). Through the above analysis it can be concluded that the origin of the low solar reflectivity of our as-prepared porous two-phase HPCs is the insufficiency of micropores, while the spectral contributions of the constituent components in the porous three-phase hybridized HPCs for the daytime radiative cooling enhancement is the added scattering interfaces between silica-microspheres and hierarchical porous coatings.

## ■ CONCLUSION

In summary, we demonstrated an efficient daytime radiative cooler with a high interface optical scattering and cooling performance achieved by enabling a three-phase self-assembled hybrid porous composite consisting of silica microspheres embedded into the hierarchical porous polymers. The flexible daytime radiative cooler exhibits an averaged temperature drop of approximately10℃ beneath the ambient, even under intense solar irradiance. The hybridization of an appropriate amount of silica microspheres provides a feasible remedy for the low solar reflectivity of non-optimized two-phase porous HPCs at near-infrared. It is realized by enhancing the reflectivity via the added silica/air interfaces while maintaining the near-perfect emissivity of primary HPCs. The MC simulation suggests that the origin of the unexpected low solar reflectivity in our as-prepared two-phase porous HPCs is ascribed to the lack of micropores and the enhancement by microspheres mainly accounts for the silica/air interfaces rather than the silica/polymer counterparts. The investigation of three-phase self-assembled hybrid porous composites as well as the cooperative contribution from both the hierarchical pores and microspheres provides an alternative to flexible daytime radiative cooling.

### ■ EXPERIMENTAL AND THEORETICAL METHODS

**Material preparation.** The suspended solution of silica microspheres with a diameter of 8 microns in 50vol% ethanol (J&K) was used as received to form a well-dispersed microsphere suspension by ultrasonicator. The suspended solution was then heated at a constant temperature to get pure silica microspheres. The silica microspheres and P(VdF-HFP) (Pellets, ALDRICH) polymer were weighted to control the final mass fractions of 3%, 6%, 9%, 12%, and 15% for the as-prepared coatings. The clean solution of acetone (AR, Chinasun), P(VdF-HFP) and de-ionized water with the mass ratio of 8:1:1was firstly formed in the glass sample vial by vigorous magnetic stirring to entirely dissolve the polymer within the mixed solvents. The silica microspheres were subsequently added and sonicated to ensure the well-dispersed suspension. Finally, the mixed solution was poured into the plastic mold with a constant thickness of 250 μm, resulting in the dry and porous coating when the solvents evaporated in the ambient atmosphere at room temperature of 25℃.

**Characterization**. The surface and cross-section morphologies of coating samples were characterized by SEM (Hitachi SU8010). The optical reflectivity of samples was measured using a UV–VIS–NIR spectrophotometer (Lambda-1050s, PerkinElmer) with a wavelength range of 300–2500 nm. The infrared emissivity over the wavelength of 5-25μm was determined by using a VERTEX 70 (Bruker) FT-IR spectrometer with an A562 integrating sphere.

**Cooling power calculation**. The net cooling power, $P_{net}$, is defined as the difference between the output and input energy with respect to a radiative cooler. The radiative cooling process involves many factors, such as solar irradiation, thermal emission, atmospheric emission, and heat conduction and convection, which are summarized in the following equation[19],

$$P_{net} = P_{rad}(T_c) - P_{sun} - P_{atm}(T_a) - P_{cc} \qquad (1)$$

where $P_{rad}(T_c) = \int \int I_{BB}(\lambda, T_c)\varepsilon(\lambda, \theta) \cos\theta \, d\Omega \, d\lambda$ means the power radiated by a cooler whose temperature is $T_c$; $P_{sun} = I_{AM1.5}(\lambda)A(\lambda) \, d\lambda$ is the absorbed power due to solar irradiation; $I_{AM1.5}(\lambda)$ is the standard solar irradiance; $A(\lambda)$ the solar absorbance; $P_{atm} = \int \int I_{BB}(\lambda, T_{amb})\varepsilon_{atm}(\lambda, \theta)\varepsilon(\lambda, \theta) \cos\theta \, d\Omega \, d\lambda$ is the absorbed power due to atmospheric opacity; $I_{BB}(\lambda, T_{amb})$ is the emissivity spectrum of blackbody at $T_{amb}$; $\varepsilon_{atm}$ and $\varepsilon$ are the emissivities of the atmosphere and radiative cooler, respectively; $P_{cc} = h_c(T_{amb} - T_c)$ is the dissipated power due to heat conduction and convection, where $h_c$ the heat transfer coefficient. In the calculation, the values $h_c$ and $T_{amb}$ are adopted to be 6.9 Wm$^{-2}$K$^{-1}$ and 313 K according to the rooftop measurement.

**Mie scattering based Monte Carlo simulation**. By using Mie theory, the scattering efficiency ($Q_{sca}$) and extinction efficiency ($Q_{ext}$) for an individual sphere can be calculated, and the absorption efficiency ($Q_{abs}$) can be obtained by taking their difference[44–47],

$$Q_{ext}(m,\chi) = \frac{2}{\chi^2}\sum_{n=1}^{\infty}(2n+1)\,\boldsymbol{Re}\{a_n + b_n\} \tag{2}$$

$$Q_{sca}(m,\chi) = \frac{2}{\chi^2}\sum_{n=1}^{\infty}(2n+1)(|a_n|^2 + |b_n|^2) \tag{3}$$

$$Q_{abs} = Q_{ext} - Q_{sca} \tag{4}$$

where $\boldsymbol{Re}$ represents the real part of a complex number. The Mie coefficients of $a_n$, $b_n$ can be obtained by,

$$a_n = \frac{\psi_n'(m\chi)\psi_n(\chi) - m\psi_n(m\chi)\psi_n'(\chi)}{\psi_n'(m\chi)\xi_n(\chi) - m\psi_n(m\chi)\xi_n'(\chi)} \tag{5}$$

$$b_n = \frac{m\psi_n'(m\chi)\psi_n(\chi) - \psi_n(m\chi)\psi_n'(\chi)}{m\psi_n'(m\chi)\xi_n(\chi) - \psi_n(m\chi)\xi_n'(\chi)} \tag{6}$$

where $\xi_n = \psi_n - i\chi_n$ and $\psi_n$, $\chi_n$ are Riccati–Bessel functions; $\chi = \frac{\pi D}{\lambda}$ is the size parameter; $m = \frac{n+ik}{n_0}$ is the normalized complex refractive index; $n_0$ and $n+ik$ are the refractive indexes of the sphere and non-absorbent medium, respectively. Another essential quantity is the asymmetric parameter, $g$, which can be obtained by,

$$\begin{aligned}g &= Q_{sca}\langle\cos\theta\rangle \\ &= \frac{4}{\chi^2}\left[\sum_{n=1}^{\infty}\frac{n(n+2)}{n+1}\boldsymbol{Re}(a_n a_{n+1}^* + b_n b_{n+1}^*) + \sum_{n=1}^{\infty}\frac{(2n+1)}{n(n+1)}\boldsymbol{Re}(a_n b_n^*)\right]\end{aligned} \tag{7}$$

where the notation $b_n^*$ represents the conjugate transpose of $b_n$. The value of $g$ is confined in the range of $[-1,1]$, which means backward (forward) scattering is dominant as $g$ is negative (positive).

After the microscopic scattering parameters of a single microsphere are obtained, they are converted into the macroscopic parameters of the microsphere-polymer composite coating by the summation principle, due to the interaction between multiple microspheres can be ignored in the sparse particle system. The effective scattering and absorption coefficients, as well as the asymmetric parameter of a macroscopic coating, can be described as,

$$\sigma = \sum_{i=1}^{c} \frac{3 f_i Q_{s,i}}{2 d_i} \tag{8}$$

$$\kappa = \sum_{i=1}^{c} \frac{3 f_i Q_{a,i}}{2 d_i} \tag{9}$$

$$g_t = \frac{1}{\sigma} \sum_{i=1}^{c} \frac{3 f_i g_i Q_{s,i}}{2 d_i} \tag{10}$$

where $c$ is the number of different microspheres used; $f_i$ is the volume concentration of each kind of microsphere. The variables obtained in Equation (8-10) can be put into the Monte Carlo program[26,48–51] to calculate the samples' measurable reflection. Our computational program is based on an open-source code written by Matzler[52], which is modified to meet the current work requirements on the aspect of the material's wavelength-dependent dispersion and multiple kinds of scatters.

■ **ASSOCIATED CONTENT**

**Supporting Information**

The Supporting Information is available free of charge at https://pubs.acs.org/doi/xxx.

Macroscopic morphologies of HPCs with different thicknesses (Figure S1), home-made template for preparing homogeneous coatings (Figure S2), SEM images of the HPCs hybridized by silica microspheres (Figure S3, Figure S4), long-time continual rooftop temperature test and the accompanying meteorological conditions (Figure S5), simulated reflection spectra using Monte Carlo method (Figure S6).

**Notes**

The authors declare no competing financial interest.

## ■ ACKNOWLEDGMENTS

This work was financially supported by the National Key Research and Development Program of China (No. 2017YFB1002900), the National Natural Science Foundation of China (Nos. 61405132, 51661145021), the Key Natural Science Program of Jiangsu Province (Nos. BK20181167, BE2016772 and BK20151540), the Opening Project of State Key Laboratory of High Performance Ceramics and Superfine Microstructure (No. SKL201912SIC), and the Traction Project of Key Laboratory of Advanced Carbon Materials and Wearable Energy Technologies of Jiangsu Province (No. Q816000217), and the Priority Academic Program Development (PAPD) of Jiangsu Higher Education Institutions.